\documentclass[11pt,a4paper]{article}
\usepackage{jheppub}
\usepackage[T1]{fontenc}
\usepackage{amssymb}
\usepackage{mathtools}
\usepackage{stackengine}
\usepackage{scalerel}

\usepackage{hyperref}
\usepackage{epsfig}
\usepackage{amsmath,latexsym,amssymb}
\usepackage{graphicx}
\usepackage{braket}
\usepackage{pdfsync}

\usepackage{framed}
\usepackage{mathtools}

\usepackage{jheppub}
\usepackage[T1]{fontenc}
\usepackage{amssymb}
\usepackage{mathtools}
\usepackage{amsmath}
\usepackage{bm}
\usepackage{stackengine}

\usepackage{scalerel}
\usepackage{eufrak}
\usepackage{framed}

\usepackage{mathtools}

\usepackage{pstricks}

\usepackage{amsmath}
\usepackage{bbold}
\usepackage{bm}
\usepackage{latexsym}
\usepackage{braket}
\usepackage{slashed}
\usepackage{graphicx,booktabs,multirow}
\usepackage{eufrak}
\numberwithin{equation}{section}

\usepackage{color}
\usepackage{pstricks}
\usepackage{amssymb}

\makeatletter
\newcommand{\doublewidetilde}[1]{{%
  \mathpalette\double@widetilde{#1}%
}}
\newcommand{\double@widetilde}[2]{%
  \sbox\z@{$\m@th#1\widetilde{#2}$}%
  \ht\z@=.9\ht\z@
  \widetilde{\box\z@}%
}
\makeatother

\usepackage{hyperref}
\usepackage{slashed}
\usepackage{epsfig}
\usepackage{amsmath,latexsym,amssymb}
\usepackage{graphicx}
\usepackage[latin1]{inputenc}
\usepackage{braket}
\usepackage{pdfsync}

\usepackage{framed}

\usepackage{mathtools}

\def\be{\begin{equation}}
\def\ee{\end{equation}}
\def\ba{\begin{eqnarray}}
\def\ea{\end{eqnarray}}

\newcommand{\bz}{\bar{z}}

\newcommand{\bh}{\bar{h}}

\newcommand{\comment}[1]{}

\newcommand{\eea}{\end{eqnarray}}

\author{
Tomasz R.\ Taylor${}^{1,2}$,\, Bin Zhu${}^3$\\[0.5cm]
 $^1${\it Department of Physics \\
  Northeastern University, Boston, MA 02115, USA}\\
  $^2${\it Faculty of Physics, University of Warsaw, ul. Pasteura 5, 02-093 Warsaw, Poland}\\
$^3${\it Perimeter Institute for Theoretical Physics,\\ Waterloo, ON N2L 2Y5, Canada}\\[0.2cm]
}

\emailAdd{taylor@neu.edu}
\emailAdd{bzhu1@perimeterinstitute.ca}

\title{\boldmath {Celestial Supersymmetry} \unboldmath}

\abstract{We discuss supersymmetric Yang-Mills theory coupled to dilatons in the framework of celestial holography. We show that in the presence of point-like dilaton sources, the CCFT operators associated with the gauge supermultiplet acquire a simple, factorized form. They factorize into the holomorphic (super)current part and the exponential ``light''  operators of Liouville theory, in the infinite central charge limit. The current sector exhibits (1,0) supersymmetry, thus implementing spacetime supersymmetry in CCFT.}

\gdef\@fpheader{}
\makeatother

\begin{document}
\maketitle
\section{Introduction}
In string theories with spacetime supersymmetry, two-dimensional conformal field theories (CFTs) describing the superstring world-sheets are also symmetric under two-dimensional superconformal transformations including two-dimensional supersymmetry. Physics in higher-dimensional ambient spacetime is described by  two-dimensional superconformal field theories. The goal of celestial holography is to describe physics in four-dimensional asymptotically flat spacetime as a hologram on two-dimensional celestial sphere, in the framework of celestial conformal field theory (CCFT) \cite{Strominger:2017zoo,Raclariu:2021zjz,Pasterski:2021rjz}. In this context, there is an obvious question: Is there any relation between four-dimensional supersymmetry and some type of supersymmetry in CCFT, hereafter called celestial supersymmetry?

The symmetries of CCFT  reflect the BMS symmetries of asymptotically flat spacetime. In
Ref.\cite{Fotopoulos:2020bqj}, we discussed  (one type of) supersymmetric extensions of BMS algebra.
We found that spacetime supersymmetry is disconnected from  two-dimensional supersymmetry.
The reason was that spacetime supersymmetry algebra includes supertranslations, which are genuinely nonholomorphic on a celestial sphere, while the two-dimensional superconformal algebras  have factorized holomorphic and antiholomorphic parts. More recently, however, the role of (super)translations came under scrutiny because the CCFT correlators obtained by taking Mellin transforms of scattering amplitudes are overconstrained by momentum conservation  \cite{Mizera:2022sln}.
One possible resolution of this problem relies on introducing background fields that violate translational invariance \cite{Fan:2022vbz,Fan:2022kpp,Casali:2022fro,PipolodeGioia:2022exe,Gonzo:2022tjm,Melton:2022fsf,Costello:2022wso,Banerjee:2023rni}. With translational symmetry broken in this way, our question can be rephrased as: Are there some background field configurations that admit celestial supersymmetry? In this work, we give an affirmative answer to this question.

In Ref.\cite{Stieberger:2022zyk}, we considered the Yang-Mills theory coupled to a complex dilaton field and introduced a point-like source creating a dilaton shockwave. This background breaks four-dimensional translational symmetry in a controllable way and supplies external momentum to the gauge system. Multi-gluon celestial MHV amplitudes evaluated in such a background have simple, factorized structure.
They factorize into the holomorphic current correlator times the correlator of a Liouville theory with an infinite central charge.\footnote{The factorizations between the current sector and the other part that decouples from the current part was suggested earlier in Ref.\cite{Nande:2017dba}.}
In the current sector, the correlators are holomorphic. They contain all the information about the spins of celestial primaries and the gauge group structures. In the Liouville sector, we encountered the correlation functions of  ``light'' operators, evaluated in the limit of the Liouville coupling $b\rightarrow 0$ (infinite central charge).

The factorization of celestial amplitudes into the current and Liouville sectors allows for addressing the question posed at the very beginning in a simpler way. We consider supersymmetric Yang-Mills theory and show that in the presence of a point-like dilaton source introduced in Ref.\cite{Stieberger:2022zyk}, the current sector exhibits (1,0) celestial supersymmetry. For comparison, in heterotic superstring theory, the left-moving current sector is not supersymmetric, while the right-moving part has a similar world-sheet supersymmetry.

Other aspects of fermionic fields and supersymmetries in CCFT have been discussed in Refs.\cite{Iacobacci:2020por,Narayanan:2020amh,Pasterski:2020pdk,Jiang:2021xzy,Brandhuber:2021nez,Hu:2021lrx,Ferro:2021dub,Himwich:2021dau,Pano:2021ewd,Jiang:2021ovh,Ahn:2021erj,Ahn:2022oor,Bu:2021avc,Ahn:2022vfw,Hu:2022bpa,Banerjee:2022lnz}.

\section{Review of (1,0)  superconformal symmetry in two dimensions}
In this section, we give a brief review of $(1,0)$ superconformal symmetry in two dimensions, by following Refs.\cite{Bershadsky:1985dq,Friedan:1984rv}. See also Ref.\cite{Polchinski:1998rr}.

There are two superconformal algebras, Neveu-Schwarz (NS) algebra and Ramond algebra, corresponding to two different periodicities of the fermionic operators. All correlators discussed in the present work
are single-valued, with the trivial monodromy when a fermionic operator circulates around the other one. For that reason, only the NS algebra is directly related to our work and we will not discuss the Ramond algebra.

The $(1,0)$ superconformal algebra is generated by the super stress-energy tensor
\begin{align}
\mathbf{T}(Z) = T_F(z) +\theta\,  T_B(z) \, ,
\end{align}
where $Z=(z,\theta)$ is the superspace coordinate, $T_B$ is the usual stress-energy tensor, and $T_F(z)$ is its superpartner with conformal weights $\Delta=h=\frac{3}{2}, \bar{h} = 0$. The OPEs among them are
\begin{align}
T_B(z_1)\, T_B(z_2) &= \frac{c}{2\, z_{12}^4} +\frac{2}{z_{12}^2}T_B(z_2) +\frac{1}{z_{12}}\partial_z T_B(z_2) \, , \label{eq:TT}\\
T_B(z_1) \, T_F(z_2) & = \frac{3}{2\, z_{12}^2}T_F(z_2) +\frac{1}{z_{12}} \partial_z T_F(z_2) \, , \label{eq:TTF}\\
T_F(z_1) \, T_F(z_2) &= \frac{2c}{3 \, z_{12}^3} + \frac{2}{z_{12}} T_B(z_2) \, , \label{eq:TFTF}
\end{align}
For  $T_B(z)$ and $T_F(z)$ the Laurent expansions are
\begin{align}
T_B(z) &= \sum_{m=-\infty}^{\infty} \frac{L_m}{z^{m+2}} \, , \\
T_F(z) &= \sum_{r\in \mathbb{Z}+ \frac{1}{2}} \frac{G_r}{z^{r+\frac{3}{2}}} \, .
\end{align}
Note that in the NS case, the sum is over half-integer $r$.
One finds the following NS algebra from the OPEs Eqs.(\ref{eq:TT})-(\ref{eq:TFTF})
\begin{align}
[L_m, L_n] &= (m-n)\, L_{m+n} +\frac{c}{12}(m^3-m)\delta_{m,-n} \, , \\
\{G_r, G_s\} &= 2 \, L_{r+s} +\frac{c}{12}(4r^2-1) \delta_{r,-s} \, , \\
[L_m, G_r] &= \frac{m-2r}{2} G_{m+r} \, .
\end{align}
The global superconformal group OSP(2|1) is generated by $L_{-1}, \, L_0, \, L_1,\, G_{-1/2}, \, G_{1/2}$. \par
By using the superfield formalism, a NS primary superfield with holomorphic conformal weight $\Delta = h$ can be represented as
\begin{align}
\Phi_{\Delta}(z,\theta) = \phi_{\Delta}(z) +\theta \, \psi_{\Delta+1/2}(z) \, .
\end{align}
The transformation properties of the NS primary superfield under $T_B$ and $T_F$ are determined by the following OPEs
\begin{align}
T_B(z_1) \, \phi(z_2) &= \frac{\Delta}{z_{12}^2} \phi(z_2) +\frac{1}{z_{12}}\partial_z \phi(z_2) \, , \\
T_F(z_1) \, \phi(z_2) &= \frac{1}{z_{12}}\psi(z_2) \, ,\\
T_F(z_1) \, \psi(z_2) &= \frac{2\Delta}{z_{12}^2} \phi(z_2) +\frac{1}{z_{12}}\partial_z \phi(z_2) \, .
\end{align}
By using the mode expansion of $T_B$ and $T_F$, one can obtain the following commutators:
\begin{align}
[L_n, \Phi(z,\theta)] &= \Big(z^{n+1}\partial_z +(n+1) z^n ( \Delta+\frac{1}{2}\theta\partial_\theta)\Big) \Phi(z,\theta) \, , \label{eq:LmPhi} \\
[G_{n+\frac{1}{2}} ,\Phi(z,\theta) ] &= \Big( z^{n+1}(\partial_\theta -\theta\partial_z) -(n+1)\, z^n 2 \, \Delta  \, \theta \Big) \, \Phi(z,\theta) \, .\label{eq:GmPhi}
\end{align}
For the global generators $L_{-1}, \, L_0, \, L_1,\, G_{-1/2}, \, G_{1/2}$, Eqs.(\ref{eq:LmPhi})-(\ref{eq:GmPhi}) give us the global superconformal Ward identities of the supercorrelators.
In particular, the Ward identities generated by $G_{-1/2}$ and $G_{1/2}$ will be used in the following sections.

\section{4D theory: $\mathcal{N}=1$ SYM coupled to a massive chiral multiplet}
We consider the following $\mathcal{N}=1$ supersymmetric theory with SYM coupled to a massive chiral dilaton supermultiplet $(\phi,\chi)$. This is a supersymmetric completion of the usual dilaton-YM Lagrangian density, see Ref.\cite{Dixon:2004za}:
\begin{align}
\mathcal{L} &= \int d^4\theta \, \Phi^\dagger \Phi +\int d^2\theta \left[ \frac{m_H}{2} \Phi^2 + \frac{1}{4}(1-\frac{4}{\Lambda} \Phi) \, \text{tr} \, W^\alpha W_{\alpha}\right] \nonumber\\
&\qquad +\int d^2\bar{\theta} \left[ \frac{m_H}{2} \Phi^{\dagger2} + \frac{1}{4}(1-\frac{4}{\Lambda} \Phi^{\dagger}) \, \text{tr} \, \overline{W}_{\dot{\alpha}} \overline{W}^{\dot{\alpha}}\right] \label{eq:Lag1}\\[2mm]
&= ~\partial_{\mu}\phi^{\dagger}\partial^{\mu}\phi -m_H^2 \phi^{\dagger}\phi-i\bar{\chi} \slashed{\partial}\chi  -\frac{m_H}{2} \chi\chi  -\frac{m_H}{2} \bar\chi\bar\chi    +\frac{1}{2} \text{tr} D^2-\frac{1}{4} \text{tr} G_{\mu\nu}G^{\mu\nu}-i\text{tr}\bar{\lambda}\slashed{D}\lambda \nonumber\\[1mm]
&\quad\quad+\Bigg\{\frac{1}{\Lambda} \Big[  2\sqrt{2} i\chi^{\alpha} \text{tr} \Big( \lambda_\alpha D-(\sigma_{\mu\nu})_{\alpha}^{~\beta}\lambda_\beta G^{\mu\nu}_{SD}\Big)  \label{eq:Lag2}\\
&\quad\quad\quad\quad\quad~ + \phi\, \text{tr}\Big( 2i\bar{\lambda}\slashed{D}\lambda+G_{SD \, \mu\nu}G^{\mu\nu}_{SD}+m_H\bar\lambda\bar \lambda\Big) \Big] + \, \text{h.}\, \text{c.} \Bigg\}~-\frac{1}{|\Lambda|^2}(\text{tr}\lambda \lambda)(\text{tr}\bar\lambda\bar \lambda)\ ,
\nonumber
\end{align}
where $\Lambda$ is a parameter with mass dimension 1. In the standard model, it is related to the VEV of the Higgs field \cite{Dixon:2004za}. Note that the dilaton is massive, hence the $U(1)$ R symmetry of massless theory is explicitly broken. Later, we will consider the massless limit ($m_H\rightarrow 0$) of the massive case.
Following Ref.\cite{Stieberger:2022zyk}, we also introduce a source term for the dilaton field:
\be
\mathcal{L}_{\cal J}=\frac{1}{\Lambda}{\cal J}^{\dagger}\phi+  \text{h.}\, \text{c.}\label{sourcet} \ee
with the mass dimension 4 source \be{\cal J}(x)=\delta^{(4)}(x)\ .\ee
We are interested in SYM amplitudes evaluated in the presence of this dilaton source. To that end, we integrate out the dilaton field. This leads to the following effective action:
\begin{align}
{\cal S} &= \frac{1}{\Lambda^2}\int d^4y\! \int d^4x\, {\cal J}(y)D(y-x)\text{tr}\Big[ 2i\bar{\lambda}\slashed{D}\lambda(x)+G_{SD \, \mu\nu}G^{\mu\nu}_{SD}(x)\nonumber\\
&\qquad\qquad+m_H\bar\lambda\bar \lambda(x)\Big] + \, \text{h.}\, \text{c.}+\dots,\label{act}
\end{align}
where $D(y-x)=-i(\Box+m^2)^{-1}$. In our case,
\be \int d^4y{\cal J}(y)D(y-x)= D(x)\ ,\ee
therefore the effect of the source is to introduce an additional, position-dependent coupling constant in the SYM sector, which violates translation symmetry.

The amplitudes involving gauge bosons and gauginos are related by supersymmetric Ward identities
\cite{Parke:1985pn}. The source term (\ref{sourcet}), however, breaks supersymmetry explicitly, therefore its supersymmetry variation leads to additional terms. In particular, the identities obtained from the holomorphic part of the transformations receive additional contributions proportional to ${\cal J}^\dagger$. Since ${\cal J}^\dagger$ couples to the R-symmetry violating term $m_H\lambda\lambda$, see Eq.(\ref{act}), the MHV amplitudes become related to the amplitudes involving pairs of gauginos carrying identical helicities. We will see later that the identities following from the holomorphic SUSY transformations are equivalent to two-dimensional NS supersymmetric Ward identities.

Another, more convenient way of studying supersymmetry relations is by considering the amplitudes involving on-shell dilatons (following from the lagrangian (\ref{eq:Lag2})), extending them off-shell, and then coupling them to the sources.
We are interested in the amplitudes with  a number of gauge particles and one  external dilaton, particularly in the amplitudes related by supersymmetry to the amplitudes with gluons in the MHV helicity configuration. We will be considering {\em partial} amplitudes associated with one particular group factor. There are \underline{two sets of such amplitudes}.

\underline{Set 1} contains the MHV gluons amplitudes with one dilaton,
\begin{align}
A_2(1^{-1},2^{-1},\phi) &= -\frac{1}{\Lambda} \langle 12\rangle^2 \, ,\label{mm1}\\
A_n(1^{-1},2^{-1},3^{+1},\dots n^{+1},\phi) &= \frac{1}{\Lambda} \frac{\langle 12\rangle^4}{\langle 12\rangle \langle 23\rangle \dots \langle n1\rangle} \, , \label{eq:MHVg}
\end{align}
and all other amplitudes that are related to Eq.(\ref{eq:MHVg}) by 4D SUSY Ward identities
\cite{Parke:1985pn,Taylor:2017sph}.
Explicit expressions for the amplitudes involving gauginos are written in Ref.\cite{Badger:2004ty}.\footnote {Supersymmetry transformations of the on-shell fields are written in the appendix of Ref.\cite{Dixon:2004za}.}
 All these amplitudes contain the same numbers of helicity ${+}1/2$ and ${-}1/2$ gauginos.


\underline{Set 2} contains the amplitudes originating from the gaugino chirality (R-symmetry) violating terms in Eq.(\ref{eq:Lag2}). They are
\begin{align}
A_2(1^{-\frac{1}{2}},2^{-\frac{1}{2}}, \phi^*) &= \frac{1}{\Lambda} m_H \, \langle 12\rangle \, , \label{mm2}\\
A_3(1^{-\frac{1}{2}},2^{-\frac{1}{2}},3^{+1},\phi^*) &= -\frac{1}{\Lambda} m_H\, \frac{ \langle 12\rangle^2}{\langle 23\rangle \langle 31\rangle} \, , \\
A_n(1^{-\frac{1}{2}},2^{-\frac{1}{2}},3^{+1}, \dots n^{+1}, \phi^*) &= -\frac{1}{\Lambda} m_H \frac{\langle 12 \rangle^3}{\langle 12\rangle \langle 23\rangle \dots \langle n1\rangle} \, . \label{eq:MHV2gim}
\end{align}
The $n$-point amplitude written  above can be derived recursively by using BCFW recursion relations \cite{Britto:2005fq}. All these amplitudes have the net surplus of two negative helicity gauginos.

It is easy to verify that Eqs.(\ref{mm1}) and (\ref{mm2}) are also valid for off-shell dilatons. BCFW recursion relations extend them to the remaining amplitudes, involving arbitrary number of on-shell gauge particles.

\section{Celestial amplitudes with a universal Liouville sector}
In this section, we construct the celestial amplitudes for the gauge particles coupled to the dilaton source (\ref{sourcet}), for set  1 and set 2 separately, such that in the limit $m_H\rightarrow 0$, the resultant celestial MHV amplitudes become related to celestial Liouville theory. In the momentum space, \be
{\cal J}(Q) = 1 \, , \label{eq:phisource}\ee
To that end, we take Mellin transforms with respect to the energies.

{}From \underline{Set 1}, we obtain
\begin{align}
\text{Celes}_1(\Delta_i|J_i)_{m_H} \supset &\,\frac{1}{\Lambda}\int \prod_{i=1}^n d\omega_i \omega_i^{\Delta_i-1} A_n(1^{-1},2^{-1},3^{+1},\dots n^{+1},\phi) \frac{1}{Q^2+m_H^2}{\cal J}(Q) \nonumber\\
&= \frac{1}{\Lambda^2}\int \prod_{i=1}^n d\omega_i \omega_i^{\Delta_i-1} \, \frac{\langle 12\rangle^4}{\langle 12\rangle \langle 23\rangle \dots \langle n1\rangle} \,  \frac{1}{Q^2+m_H^2} \, , \label{eq:Celes1massive}
\end{align}
where $\text{Celes}_1(\Delta_i|J_i)_{m_H} $ denote celestial amplitudes from  set 1, with conformal dimensions $\Delta_i$ and helicities $J_i$.
Note that in the massless dilaton limit, Eq.(\ref{eq:Celes1massive}) goes back to the amplitudes studied   in Ref.\cite{Stieberger:2022zyk}:
\begin{align}
\text{Celes}_1(\Delta_i|J_i) = \lim_{m_H\rightarrow 0} \text{Celes}_1(\Delta_i|J_i)_{m_H} \supset \frac{1}{\Lambda^2}\int \prod_{i=1}^n d\omega_i \omega_i^{\Delta_i-1} \, \frac{\langle 12\rangle^4}{\langle 12\rangle \langle 23\rangle \dots \langle n1\rangle} \,  \frac{1}{Q^2} \, . \label{eq:Celes1massless}
\end{align}

From \underline{Set 2}, we obtain
\begin{align}
\text{Celes}_2(\Delta_i|J_i)_{m_H}  \supset & \,\frac{1}{\Lambda}\int \prod_{i=1}^n d\omega_i\, \omega_i^{\Delta_i-1} A_n(1^{-\frac{1}{2}},2^{-\frac{1}{2}},3^{+1}, \dots n^{+1}, \phi^*)\frac{1}{Q^2+m_H^2}{\cal J}(Q) \nonumber\\
&= -\frac{m_H}{\Lambda^2}\int \prod_{i=1}^n d\omega_i\, \omega_i^{\Delta_i-1} \frac{ \langle 12 \rangle^3}{\langle 12\rangle \langle 23\rangle \dots \langle n1\rangle}\, \frac{1}{Q^2+m_H^2} . \label{eq:Celes2massive}
\end{align}
This amplitude vanishes in the limit of $m_H \rightarrow 0$. For the reasons that will become clear below, we are interested, however, in the limit
\begin{align}
\text{Celes}_2(\Delta_i|J_i) = \lim_{m_H\rightarrow 0}-\frac{2}{m_H} \text{Celes}_2(\Delta_i|J_i)_{m_H} = \frac{1}{\Lambda^2} \int \prod_{i=1}^n d\omega_i\, \omega_i^{\Delta_i-1} \,   \frac{ 2\, \langle 12 \rangle^3}{\langle 12\rangle \langle 23\rangle \dots \langle n1\rangle}\, \frac{1}{Q^2}  \, . \label{eq:Celes2massless}
\end{align}

As we will see below, upon the following choice of the conformal dimensions:
\begin{equation}
\left\{
\begin{array}{ll}
&  -1  \, \text{helicity gluon: \, } \Delta = i\lambda  \, , \\
&        -\frac{1}{2} \,  \text{helicity gaugino: \, } \Delta = \frac{1}{2} +i\lambda \, \\
&        +\frac{1}{2} \,  \text{helicity gaugino: \, } \Delta = \frac{1}{2} +i\lambda \,  \\
&+1  \, \text{helicity gluon: \, } \Delta = 1+i\lambda \, ,
\end{array}
\right. \label{eq:Deltas}
\end{equation}
two sets of celestial amplitudes $\text{Celes}_1$ and $\text{Celes}_2$, {\em c.f.} Eqs.(\ref{eq:Celes1massless}) and (\ref{eq:Celes2massless}), contain identical Mellin transforms, which means that they have a universal Liouville sector.
The information about the spins of the celestial primaries and the gauge group structures is contained entirely in the current sector.\par

The correspondence between celestial primaries associated to the gauge supermultiplet and the CCFT operators can be summarized as
\begin{align}
O^{a}_{\Delta,J}(z,\bar{z}) = \, O_J^a(z)\, \Gamma(\Delta-J)  \, e^{(\Delta-J)b\phi(z,\bar{z})} \, \label{eq:celetoLiouville}
\end{align}
with the implied limit of $b\rightarrow0$. In Eq.(\ref{eq:celetoLiouville}), the choice of the conformal dimensions are given by Eq.(\ref{eq:Deltas}),  and the current sector operators $O_J^a$ are
\begin{align}
O_{-1}^a(z) &= \sqrt{\frac{1}{\Lambda}}\,\widehat{j}^a(z),\label{norm1} \\ O_{-\frac{1}{2}}^a(z) &= i\sqrt{\frac{m_H}{2}}\psi^{-a}(z)  ,
 \\
O_{+\frac{1}{2}}^a(z) &=-i\sqrt{\frac{2}{m_H\Lambda}}\psi^{+a}(z), \\[2mm]  O_{+1}^a(z) &= j^a(z) \, .\label{norm2}
\end{align}
The role of the normalization factors will become clear in the next section.
Note that the operators $O_J^a$ are purely holomorphic with conformal weights $h=J$. With the choice given by Eq.(\ref{eq:Deltas}), celestial primaries with negative helicities have the same Liouville part,
\begin{align}  \Gamma(\Delta-J)  \, e^{(\Delta-J)b\phi(z,\bar{z})}=
\Gamma(1+i\lambda)e^{(1+i\lambda)b\phi(z,\bar{z})} \, ,
\end{align}
while for positive helicities
\begin{align} \Gamma(\Delta-J)  \, e^{(\Delta-J)b\phi(z,\bar{z})}=
\Gamma(i\lambda) e^{ i\lambda b \phi(z,\bar{z})} \, .
\end{align}
Note that only the positive helicity (self-dual) operators survive in the limit of $\Lambda\to\infty,\ m_H\to 0$ with $m_H\Lambda$ fixed.

\subsection{Three points}
We begin by writing explicit expressions for three-particle amplitudes. \underline{Set 1} contains
\cite{Badger:2004ty}: \\
\begin{align}
A_3(1^{-1},2^{-1},3^{+1},\phi) &= \frac{1}{\Lambda}\frac{\langle 12\rangle^3}{\langle 23 \rangle \langle 31\rangle} = \frac{1}{\Lambda}\frac{z_{12}^3}{z_{23}z_{31}} \frac{\omega_1\omega_2}{\omega_3} \, , \\
A_3(1^{-1}, 2^{-\frac{1}{2}}, 3^{+\frac{1}{2}} , \phi) &=\frac{1}{\Lambda} \frac{\langle 12\rangle^3 \langle 13 \rangle}{\langle 12\rangle \langle 23\rangle\langle31\rangle} = \label{h22} \frac{1}{\Lambda}\frac{z_{12}^2}{z_{32}} \frac{\omega_1\omega_2^{1/2}}{\omega_3^{1/2}} \, , \\
A_3(1^{-\frac{1}{2}}, 2^{-1}, 3^{+\frac{1}{2}}, \phi)&= \frac{1}{\Lambda}\frac{\langle 21\rangle^3 \langle 23\rangle}{\langle 12\rangle \langle 23\rangle \langle 31 \rangle} = \frac{1}{\Lambda}\frac{z_{12}^2}{z_{13}} \frac{\omega_1^{1/2}\omega_2}{\omega_3^{1/2}} \, ,\label{h33}
\end{align}
In \underline{set 2},
\begin{align}
A_3(1^{-\frac{1}{2}},2^{-\frac{1}{2}}, 3^{+1},\phi^*) & = -\frac{m_H}{\Lambda}\frac{\langle 12\rangle^2}{\langle 23\rangle \langle 31\rangle} =- \frac{m_H}{\Lambda}\frac{ z_{12}^2}{z_{23}z_{31}} \frac{\omega_1^{1/2}\omega_2^{1/2}}{\omega_3} \, .\label{h44}
\end{align}
By using Eqs.(\ref{eq:Celes1massless}), (\ref{eq:Celes2massless}), and (\ref{eq:Deltas}), we find the following $m_H\to 0$ limits of three-point celestial amplitudes:
\begin{align}
\text{Celes}_1(&i\lambda_1,i\lambda_2, 1+i\lambda_3|J_1=-1, \, J_2 = -1, \, J_3 = +1) \nonumber\\
=& ~\frac{1}{\Lambda^2}\frac{z_{12}^3}{z_{23}z_{31}} \int \prod_{i=1}^3 d\omega_i \omega_1^{i\lambda_1-1}\omega_2^{i\lambda_2-1}\omega_3^{i\lambda_3} \frac{\omega_1\omega_2}{\omega_3} \frac{1}{Q^2} ~=~ \frac{1}{\Lambda}
\langle \widehat{j}_1\, \widehat{j}_2 \, j_3\rangle\, L_3(z_i,\bz_i) \label{eq:C3point1}
\end{align}
and similar expressions following from Eqs.(\ref{h22}) and (\ref{h33}). Here, the universal Liouville factor \cite{Stieberger:2022zyk}
\be L_3(z_i,\bz_i)=
 {2\pi\over\Lambda} \delta\left(\sum_{i=1}^3 \lambda_i\right) \Gamma(-i\lambda_1)\Gamma(-i\lambda_2)\Gamma(1-i\lambda_3) (z_{12}\bar{z}_{12})^{i\lambda_3-1}(z_{23}\bar{z}_{23})^{i\lambda_1} (z_{13}\bar{z}_{13})^{i\lambda_2}  \, . \label{eq:C3point11}
\ee
Furthermore, by taking the limit (\ref{eq:Celes2massless}) of Eq.(\ref{h44}), we obtain
\begin{align}
\text{Celes}_2 &\left(\frac{1}{2}+i\lambda_1, \, \frac{1}{2}+i\lambda_2, \, 1+i\lambda_3| J_1=-\frac{1}{2}, \, J_2 = -\frac{1}{2}, \, J_3 = +1 \right) \nonumber\\
=& \frac{1}{\Lambda}\frac{2\, z_{12}^2}{z_{23}z_{31}} \int \prod_{i=1}^3 d\omega_i \omega_1^{i\lambda_1-1/2}\omega_2^{i\lambda_2-1/2} \omega_3^{i\lambda_3} \, \frac{\omega_1^{1/2}\omega_2^{1/2}}{\omega_3} \frac{1}{Q^2} = \langle \psi_1^- \, \psi_2^- \, j_3\rangle\,
L_3(z_i,\bz_i) \, .\label{eq:C3point4}
\end{align}
All correlators (\ref{eq:C3point1})-(\ref{eq:C3point4}) have the same three-point Liouville part. The respective current correlators are
\begin{align}
&\langle \widehat{j}_1 \, \widehat{j}_2\, j_3\rangle = \frac{z_{12}^3}{z_{23}z_{31}} \, , \label{eq:3ptmatter1}\\
&\langle \widehat{j}_1 \, \psi_2^- \,  \psi_3^+\rangle  =  \frac{z_{12}^2}{z_{32}} \, ,\label{eq:3ptmatter2}\\
&\langle \psi_1^- \, \widehat{j}_2 \, \psi_3^+\rangle  = \frac{z_{12}^2}{z_{13}} \, , \label{eq:3ptmatter3}\\
 &\langle\psi_1^-\, \psi_2^- \, j_3\rangle   = \frac{2 \,  z_{12}^2}{z_{23}z_{31}} \, . \label{eq:3ptmatter4}
\end{align}
The current operators are purely holomorphic $(\bh=0)$, with
the chiral weights equal to the helicities of the corresponding particles:
\begin{equation}
h(\widehat{j})=-1 \, , \quad h(\psi^-)=  -\frac{1}{2}, \quad h(\psi^+)= +\frac{1}{2}, \quad h(j)= +1 \, . \label{eq:conwh}
\end{equation}

\subsection{Four points}
We can perform similar computations of the four-point correlators. The relevant amplitudes are \cite{Badger:2004ty}:
\begin{align}
A_4(1^{-1},2^{-1},3^{+1},4^{+1},\phi) &= \frac{1}{\Lambda}\frac{\langle 12\rangle^4}{\langle12\rangle\langle 23\rangle\langle 34\rangle \langle 41\rangle} =\frac{1}{\Lambda} \frac{z_{12}^3}{z_{23}z_{34}z_{41}}\frac{\omega_1\omega_2}{\omega_3\omega_4} \, , \label{eq:psp4pt1}\\
A_4(1^{-1},2^{-\frac{1}{2}},3^{+\frac{1}{2}},4^{+1},\phi) &= \frac{1}{\Lambda}\frac{\langle 12\rangle^2 \langle 13\rangle}{\langle 23\rangle\langle 34\rangle\langle 41\rangle} = \frac{1}{\Lambda}\frac{z_{12}^2 z_{13}}{z_{23}z_{34}z_{41}}\frac{\omega_1\omega_2^{1/2}}{\omega_3^{1/2}\omega_4} \, , \label{eq:psp4pt2}\\
A_4(1^{-1},2^{-\frac{1}{2}},3^{+1},4^{+\frac{1}{2}},\phi) &= -\frac{1}{\Lambda}\frac{\langle 12\rangle^2}{\langle 23\rangle \langle 34\rangle} = - \frac{1}{\Lambda}\frac{z_{12}^2}{z_{23}z_{34}}\frac{\omega_1\omega_2^{1/2}}{\omega_3\omega_4^{1/2}} \, , \label{eq:psp4pt3} \\
A_4(1^{-\frac{1}{2}},2^{-1},3^{+\frac{1}{2}},4^{+1},\phi) &=\frac{1}{\Lambda} \frac{\langle 12\rangle^2}{\langle 34\rangle\langle 14\rangle} = \frac{1}{\Lambda}\frac{z_{12}^2}{z_{34}z_{14}} \frac{\omega_1^{1/2}\omega_2}{\omega_3^{1/2}\omega_4} \, ,\label{eq:psp4pt4}
\end{align}
\begin{align}
 A_4(1^{-\frac{1}{2}},2^{-1},3^{+1},4^{+\frac{1}{2}},\phi) &= \frac{1}{\Lambda}\frac{\langle 12\rangle^2 \langle 24\rangle}{\langle 23 \rangle \langle 34\rangle\langle14\rangle}= \frac{1}{\Lambda}\frac{z_{12}^2z_{24}}{z_{23}z_{34}z_{14}} \frac{\omega_1^{1/2}\omega_2}{\omega_3\omega_4^{1/2}} \, , \label{eq:psp4pt5}\\
A_4(1^{-\frac{1}{2}},2^{-\frac{1}{2}},3^{+\frac{1}{2}}, 4^{+\frac{1}{2}},\phi) &= \frac{1}{\Lambda}\frac{\langle 12\rangle^2}{\langle 23\rangle\langle 14\rangle} = \frac{1}{\Lambda} \frac{z_{12}^2}{z_{23}z_{14}} \frac{\omega_1^{1/2}\omega_2^{1/2}}{\omega_3^{1/2}\omega_4^{1/2}}\, ,\label{eq:psp4pt6}
\end{align}
from set 1 and
\begin{align}
A_4(1^{-\frac{1}{2}},2^{-\frac{1}{2}}, 3^{+1}, 4^{+1},\phi^*) = -\frac{m_H}{\Lambda}\frac{ \langle 12\rangle^3}{\langle 12\rangle \langle 23\rangle \langle 34\rangle \langle 41\rangle} = -\frac{m_H}{\Lambda} \frac{ z_{12}^2}{z_{23}z_{34}z_{41}} \frac{\omega_1^{1/2}\omega_2^{1/2}}{\omega_3\omega_4} \, \label{eq:psp4pt7}
\end{align}
from set 2.
By using Eqs.(\ref{eq:Celes1massless}), (\ref{eq:Celes2massless}), and (\ref{eq:Deltas}), we find
\begin{align}
\text{Celes}_1(&i\lambda_1,i\lambda_2, 1+i\lambda_3, 1+i\lambda_4|J_1=-1, \, J_2 = -1, \, J_3 = +1 \, J_4 =+1) \nonumber\\
=&\frac{1}{\Lambda^2}\frac{z_{12}^3}{z_{23}z_{34}z_{41}} \int \prod_{i=1}^4 d\omega_i \omega_1^{i\lambda_1}\omega_2^{i\lambda_2}\omega_3^{i\lambda_3-1}\omega_4^{i\lambda_4-1}\frac{1}{Q^2} = \frac{1}{\Lambda}\langle \widehat{j}_1 \, \widehat{j}_2\, j_3 \, j_4\rangle L_4(z_i,\bar{z}_i),
\end{align}
and similar expressions following from Eqs.(\ref{eq:psp4pt2}-\ref{eq:psp4pt6}). The current parts of these correlators are different, but they all contain the universal Liouville factor
\begin{align}
L_4(z_i,\bar{z}_i) = \frac{2\pi}{\Lambda}  \delta\left(\sum_{i=1}^4 \lambda_i\right) \Gamma(1+i\lambda_1)\Gamma(1+i\lambda_2)\Gamma(i\lambda_3)\Gamma(i\lambda_4) I_4 (z_i,\bz_i)\, ,
\end{align}
where the function $I_4(z_i,\bz_i)$ is written explicitly in Ref.\cite{Stieberger:2022zyk}. Furthermore,
\begin{align}
\text{Celes}_2&\left( \frac{1}{2} +i\lambda_1, \, \frac{1}{2}+i\lambda_2, \, 1+i\lambda_3, \, 1+i\lambda_4 : J_1 = -\frac{1}{2}, \, J_2 = -\frac{1}{2}, \, J_3 +1, \, J_4+1\right) \nonumber\\
=&~~\langle \psi_1^- \, \psi_2^- \, j_3 \, j_4\rangle L_4(z_i,\bar{z}_i) \, .
\end{align}
In this way, we obtain
\begin{align}
&\langle \widehat{j}_1 \, \widehat{j}_2\, j_3 \, j_4\rangle = \frac{z_{12}^3}{z_{23}z_{34} z_{41}} \, , \label{eq:4ptmatter1}\\
&\langle \widehat{j}_1 \, \psi_2^- \, \psi_3^+\, j_4\rangle = \frac{z_{12}^2 z_{13}}{z_{23}z_{34}z_{41}} \, ,\label{eq:4ptmatter2}\\
&\langle \widehat{j}_1 \, \psi_2^- \, j_3 \, \psi_4^+\rangle = -\frac{z_{12}^2}{z_{23}z_{34}} \, , \label{eq:4ptmatter3}\\
&\langle  \psi_1^- \, \widehat{j}_2 \, \psi_3^+ \, j_4\rangle = \frac{z_{12}^2}{z_{34}z_{14}} \, , \label{eq:4ptmatter4}\\
&\langle  \psi_1^- \, \widehat{j}_2 \, j_3 \, \psi_4^+\rangle =  \frac{z_{12}^2 z_{24}}{z_{23}z_{34} z_{14}} \, , \label{eq:4ptmatter5}\\
&\langle \psi_1^-\psi_2^-\psi_3^+\psi_4^+\rangle =  \frac{z_{12}^2}{z_{23}z_{14}} \, ,\label{eq:4ptmatter6}\\
&\langle \psi_1^- \, \psi_2^-\, j_3\, j_4\rangle =\frac{2\, z_{12}^2}{z_{23}z_{34}z_{41}} \, .\label{eq:4ptmatter7}
\end{align}

\subsection{Celestial OPEs from the celestial current algebra and Liouville CFT}
The OPEs of the CCFT operators
(\ref{eq:celetoLiouville}) can be computed by using the OPEs of current operators and the well-known OPEs of (light) Liouville operators \cite{ZZlecture}. We want to compare them with the OPEs extracted from the collinear limits of celestial amplitudes \cite{Fotopoulos:2020bqj,Fan:2019emx,Pate:2019lpp}.
\paragraph{gluon-gluon:}
The current-current OPE is
\be
j^a(z_1) j^b(z_2) \sim \frac{f^{abc}}{z_{12}} j^c(z_2) \, .\ee
The Liouville OPE is
\begin{align} \Gamma(\Delta_1-1)&e^{(\Delta_1-1)b\phi(z_1,\bar{z}_1)} \Gamma(\Delta_2-1)e^{(\Delta_2-1)b\phi(z_2,\bar{z}_2)} \nonumber\\ &
= B(\Delta_1-1,\Delta_2-1) \Gamma(\Delta_1+\Delta_2-2) e^{(\Delta_1+\Delta_2-2)b\phi(z_2,\bar{z}_2)} \, . \label{eq:LiouvilleOPE1}
\end{align}
As a result,
\begin{align}
O^a_{\Delta_1, +1}(z_1,\bar{z}_1) O^b_{\Delta_2, +1}(z_2,\bar{z}_2) \sim B(\Delta_1-1,\Delta_2-1) \frac{f^{abc}}{z_{12}} O^c_{\Delta_1+\Delta_2-1,+1}(z_2,\bar{z}_2) \, ,
\end{align}
in agreement with Refs.\cite{Fan:2019emx,Pate:2019lpp}. \par
The current-current OPE with opposite helicities is
\be
j^a(z_1) \widehat{j}^b(z_2) \sim \frac{f^{abc}}{z_{12}} \widehat{j}^c(z_2) \, .\ee
The Liouville OPE is
\begin{align} \Gamma(\Delta_1-1)&e^{(\Delta_1-1)b\phi(z_1,\bar{z}_1)} \Gamma(\Delta_2+1)e^{(\Delta_2+1)b\phi(z_2,\bar{z}_2)} \nonumber\\ &
= B(\Delta_1-1,\Delta_2+1) \Gamma(\Delta_1+\Delta_2) e^{(\Delta_1+\Delta_2)b\phi(z_2,\bar{z}_2)} \, . \label{eq:LiouvilleOPE1a}
\end{align}
As a result,
\begin{align}
O^a_{\Delta_1, +1}(z_1,\bar{z}_1) O^b_{\Delta_2, -1}(z_2,\bar{z}_2) \sim B(\Delta_1-1,\Delta_2+1) \frac{f^{abc}}{z_{12}} O^c_{\Delta_1+\Delta_2-1,-1}(z_2,\bar{z}_2) \, ,
\end{align}
in agreement with Refs.\cite{Fan:2019emx,Pate:2019lpp}.\footnote{The antiholomorphic terms do not appear in these OPEs because the
background dilaton field couples to the MHV sector only. This was explained in detail in
Ref.\cite{Fan:2022vbz}, where the absence of antiholomorphic terms was attributed to the ``MHV projection.''}

\paragraph{gluon-gluino:}
The current-supercurrent OPE is
\be
 j^a(z_1) \psi^{+b}(z_2) \sim \frac{f^{abc}}{z_{12}} \psi^{+c}(z_2)  \, .\ee
 The Liouville OPE is \begin{align}
\Gamma(\Delta_1-1)&e^{(\Delta_1-1)b\phi(z_1,\bar{z}_1)} \Gamma\Big(\Delta_2-\frac{1}{2}\Big) e^{(\Delta_2-1/2)b\phi(z_2,\bar{z}_2)}\nonumber\\
 =& ~B\Big(\Delta_1-1, \Delta_2-\frac{1}{2}\Big) \Gamma\Big(\Delta_1+\Delta_2-\frac{3}{2}\Big) e^{(\Delta_1+\Delta_2-3/2)b\phi(z_2,\bar{z}_2)} \, . \label{eq:LiouvilleOPE2}
\end{align}
As a result,
\begin{align}
O^a_{\Delta_1,+1}(z_1,\bar{z}_1) O^b_{\Delta_2, +\frac{1}{2}}(z_2,\bar{z}_2) \sim B\Big(\Delta_1-1,\Delta_2-\frac{1}{2}\Big) \frac{f^{abc}}{z_{12}} O^c_{\Delta_1+\Delta_2-1, +\frac{1}{2}}(z_2,\bar{z}_2) \, ,
\end{align}
in agreement with   Ref.\cite{Fotopoulos:2020bqj}.\par
The current-supercurrent OPE with opposite helicities is
\be
 j^a(z_1) \psi^{-b}(z_2) \sim \frac{f^{abc}}{z_{12}} \psi^{-c}(z_2)  \, .\ee
 The Liouville OPE is \begin{align}
\Gamma(\Delta_1-1)&e^{(\Delta_1-1)b\phi(z_1,\bar{z}_1)} \Gamma\Big(\Delta_2+\frac{1}{2}\Big) e^{(\Delta_2+1/2)b\phi(z_2,\bar{z}_2)}\nonumber\\
 =& ~B\Big(\Delta_1-1, \Delta_2+\frac{1}{2}\Big) \Gamma\Big(\Delta_1+\Delta_2-\frac{1}{2}\Big) e^{(\Delta_1+\Delta_2-1/2)b\phi(z_2,\bar{z}_2)} \, .
\end{align}
As a result,
\begin{align}
O^a_{\Delta_1,+1}(z_1,\bar{z}_1) O^b_{\Delta_2, -\frac{1}{2}}(z_2,\bar{z}_2) \sim B\Big(\Delta_1-1,\Delta_2+\frac{1}{2}\Big) \frac{f^{abc}}{z_{12}} O^c_{\Delta_1+\Delta_2-1, -\frac{1}{2}}(z_2,\bar{z}_2) \, ,
\end{align}
in agreement with   Ref.\cite{Fotopoulos:2020bqj}.\par

\paragraph{gluino-gluino:}
The supercurrent-supercurrent OPE is
\be
\psi^{-a}(z_1) \psi^{+b}(z_2) \sim~ \frac{f^{abc}}{z_{12}} \, \widehat{j}^c(z_2) .\ee
The Liouville OPE is \begin{align}\Gamma\Big(\Delta_1+\frac{1}{2}\Big) \, &e^{(\Delta_1 +1/2)b\phi(z_1,\bar{z}_1)} \Gamma\Big(\Delta_2-\frac{1}{2}\Big) \, e^{(\Delta_2 -1/2)b\phi(z_2,\bar{z}_2)} \nonumber\\
=& ~B\Big(\Delta_1+\frac{1}{2}, \Delta_2 -\frac{1}{2} \Big) \Gamma(\Delta_1+\Delta_2) e^{(\Delta_1+\Delta_2)b\phi(z_2,\bar{z}_2)} \, ,
\end{align}
As a result,
\begin{align}
{O}^{a}_{\Delta_1,-\frac{1}{2}}(z_1,\bar{z}_1){O}^{b}_{\Delta_2,+\frac{1}{2}}(z_2,\bar{z}_2) \sim
B\left( \Delta_1+\frac{1}{2},\Delta_2-\frac{1}{2}\right)
\frac{f^{abc}}{z_{12}} {O}^c_{\Delta_1+\Delta_2-1,-1}(z_2,\bar{z}_2) \, , \label{eq:gigiOPE}\end{align}
in agreement with  Ref.\cite{Fotopoulos:2020bqj}.

\section{Current correlators in (1,0) superspace}
In this section, we assemble  the current and supercurrent operators into the multiplets of (1,0) supersymmetry. This is most succinctly done in (1,0) superspace parametrized by $(z, \theta)$,
where the global supersymmetry generators act on the primary superfields  in the following way:
\begin{align}G_{-1/2}\Phi_\Delta(z,\theta)&= (\partial_\theta-\theta \partial_z)\,\Phi_\Delta(z,\theta)\, ,\\
G_{+1/2}
\Phi_\Delta(z,\theta)&= [z(\partial_{\theta}-\theta\partial_z) -2 \Delta \, \theta]\,\Phi_\Delta(z,\theta) \, .
\end{align}
We introduce the following supercurrent superfields:
\begin{align}
\mathbf{\widehat J}^a(z,\theta) &= \widehat{j}^a(z)+\theta \, \psi^{-a}(z) \, ,\label{eq:bfO-}\\
\mathbf{J}^a(z,\theta) &= -\psi^{+a}(z) +\theta \,  j^a(z) \, , \label{eq:bfO+}
\end{align}
with dimensions $\Delta_{\mathbf{\widehat J}}=-1$ and $\Delta_{\mathbf{J}}=1/2$, respectively.
For the moment, we skip the gauge group indices and focus on the partial correlators associated with a single gauge group factor.

\subsection{Three points}
All nonvanishing three-point current correlators are written in
Eqs.(\ref{eq:3ptmatter1})-(\ref{eq:3ptmatter4}). We can assemble them into
\begin{align}
\langle\mathbf{\widehat J}_1\mathbf{\widehat J}_2\mathbf{J}_3
\rangle
=& ~\theta_3\langle \widehat{j}_1 \, \widehat{j}_2 \, j_3\rangle-\theta_2 \langle \widehat{j}_1 \, \psi_2^- \, \psi_3^+\rangle-\theta_1 \langle \psi_1^- \, \widehat{j}_2 \, \psi_3^+\rangle-\theta_1\theta_2\theta_3\langle \psi_1^-\, \psi_2^- \, j_3\rangle \nonumber\\
=&~ \theta_3 \frac{z_{12}^3}{z_{23}z_{31}} -\theta_2 \frac{z_{12}^2}{z_{32}} -\theta_1 \frac{z_{12}^2}{z_{13}} -\theta_1\theta_2\theta_3 \frac{2 \, z_{12}^2}{z_{23}z_{31}} \, . \label{eq:3ptO-O-O+}
\end{align}
It is easy to check that they satisfy the Ward identities associated with
$G_{-1/2}$ and $G_{+1/2}$:
\begin{align} \sum_{i=1}^3 &(\partial_{\theta_i}-\theta_i\partial_{z_i})\,
\langle\mathbf{\widehat J}_1\mathbf{\widehat J}_2\mathbf{J}_3
\rangle
 = 0 \, ,\\
\sum_{i=1}^3 &[z_i(\partial_{\theta_i}-\theta_i\partial_{z_i}) -2 \Delta_i  \theta_i]\,
 \langle\mathbf{\widehat J}_1\mathbf{\widehat J}_2\mathbf{J}_3
\rangle  = 0\, .
\end{align}

The correlator (\ref{eq:3ptO-O-O+}) also has a compact expression in terms of superspace coordinates $Z=(z,\theta)$. We use the conventions in Refs.\cite{DiVecchia:1984nyg,Fuchs:1986ew}, where the intervals in superspace are denoted as
\begin{align}
\theta_{ij} = \theta_i-\theta_j \, ,\\
Z_{ij} = z_{ij}-\theta_i\theta_j \, .
\end{align}
A generic three-point super-correlator can be written as
\begin{align}
\langle \Phi_1(Z_1) \Phi_2(Z_2)\Phi_3(Z_3)\rangle = \prod_{i<j}^3 \frac{1}{Z_{ij}^{\Delta_{ij}}} (c_{123}+c'_{123}\theta_{123}) \, , \label{eq:3ptsucogeneral}
\end{align}
where
\begin{align}
\Delta_{ij} &= \Delta_i+\Delta_j-\epsilon_{ijk}\Delta_k \, ,\\
\theta_{ijk} &= \frac{1}{\sqrt{Z_{ij}Z_{jk}Z_{ki}}} (\theta_i Z_{jk} +\theta_j Z_{ki} +\theta_k Z_{ij} +\theta_i\theta_j\theta_k) \, ,
\end{align}
and $c_{123}$, $c'_{123}$ are independent three-point coefficients. \par
The correlator (\ref{eq:3ptO-O-O+}) can be written as
\begin{align} \langle\mathbf{\widehat J}_1\mathbf{\widehat J}_2\mathbf{J}_3
\rangle&= \frac{Z_{12}^2}{Z_{23} Z_{31}} (\theta_1 Z_{23} +\theta_2 Z_{31} +\theta_3 Z_{12} +\theta_1\theta_2\theta_3) \nonumber\\
&= \frac{Z_{12}^{5/2}}{Z_{23}^{1/2}Z_{31}^{1/2}} \theta_{123} \, , \label{eq:3ptsucorr}
\end{align}
which takes the same form as Eq.(\ref{eq:3ptsucogeneral}), with
\begin{align}
c_{123}=0, \quad c'_{123} = 1 \, .
\end{align}

\subsection{Four points and OPEs}
All non-vanishing four-point current correlators are written in Eqs.(\ref{eq:4ptmatter1})-(\ref{eq:4ptmatter7}). Other correlators vanish for various reasons.
For example, $\langle \widehat{j}_1 \, \widehat{j}_2 \, \psi_3^+ \, \psi_4^+\rangle $, which receives contributions from the background due to the source ${\cal J}_1(Q,m_H)_\phi=1/\Lambda$, is of order $m_H$ and vanishes in the $m_H=0$ limit. The nonvanishing correlators can be assembled into
\begin{align}
\langle\mathbf{\widehat J}_1\mathbf{\widehat J}_2\mathbf{J}_3\mathbf{J}_4
\rangle
=&~\theta_3\theta_4 \frac{z_{12}^3}{z_{23}z_{34}z_{41}} \, - \,  \theta_2\theta_4 \frac{z_{12}^2 z_{13}}{z_{23}z_{34}z_{41}} -\theta_2\theta_3 \frac{z_{12}^2}{z_{23}z_{34}}-\theta_1\theta_4 \frac{z_{12}^2}{z_{34}z_{14}} \nonumber\\
&+\theta_1\theta_3 \frac{z_{12}^2 z_{24}}{z_{23}z_{34}z_{14}} -\theta_1\theta_2 \frac{z_{12}^2}{z_{23}z_{14}} -\theta_1\theta_2\theta_3\theta_4 \frac{2\, z_{12}^2}{z_{23}z_{34}z_{41}} \, . \label{eq:4ptO-O-O+O+}
\end{align}
Once again, one can check that the correlator (\ref{eq:4ptO-O-O+O+}) satisfies two-dimensional supersymmetric Ward identities:
\begin{align}
 \sum_{i=1}^4 &(\partial_{\theta_i}-\theta_i\partial_{z_i})\,
  \langle\mathbf{\widehat J}_1\mathbf{\widehat J}_2\mathbf{J}_3\mathbf{J}_4
\rangle = 0  \, ,\\
\sum_{i=1}^4  & [ z_i(\partial_{\theta_i}-\theta_i \partial_{z_i}) -2 \Delta_i \theta_i]\,
\langle\mathbf{\widehat J}_1\mathbf{\widehat J}_2\mathbf{J}_3\mathbf{J}_4
\rangle
 = 0 \, .
\end{align}

In the previous section, we wrote the OPEs of the current operators. In superspace, they read
\begin{align}
\mathbf{ J}^{a}(Z_1) \mathbf{J}^b(Z_2) \sim \frac{f^{abc}\, \theta_{ij}}{Z_{12}}\,  \mathbf{ J}^c(Z_2) \, ,\label{eq:superJJOPE} \\
\mathbf{ J}^{a}(Z_1) \mathbf{\widehat J}^b(Z_2) \sim \frac{f^{abc}\, \theta_{ij}}{Z_{12}} \, \mathbf{\widehat J}^c(Z_2) \, ,\label{eq:superJJhatOPE}
\end{align}
Note that Eq.(\ref{eq:superJJOPE}) is the standard OPE of the super WZW model with level $k=0$ \cite{DiVecchia:1984nyg,Fuchs:1986ew}.
\section{Relation between spacetime and celestial supersymmetries}
In order to exhibit celestial supersymmetry, it was necessary to combine two sets of amplitudes, number 1 and 2, described in section 4.  In the massless dilaton limit, the amplitudes of set 2 vanish while the amplitudes of set 1 are related by standard (spacetime) supersymmetric Ward identities. In this section, we show that when the dilaton is massive, both sets are related by spacetime supersymmetry.
With the celestial operators normalized as in Eqs.(\ref{norm1})-(\ref{norm2}), and with the massless limit taken as in section 4, this implies the celestial supersymmetry relations described in section 5.

In the first step, we want to find a supersymmetric Ward identity for the amplitude
$A_3(1^{-1/2},2^{-1/2}, 3^{+1}, \phi^*)$ from set 2, with the massive dilaton $\phi^*$ carrying momentum $P$, $P^2=m_H^2$. Following Ref.\cite{Schwinn:2006ca}, we introduce an arbitrary lightlike reference vector $r$ and define the lightlike momentum
\be
k_4^\mu\equiv P^\mu-\frac{m_H^2}{2 (P\cdot r)} r^\mu\, .  \ee
Note that
\be 2(P\cdot r)= 2(k_4\cdot r)=\langle 4r\rangle [r4]\ .\ee
The vector $r$ defines the dilatino spin quantization axis and enters into the dilatino  wave functions \cite{Schwinn:2006ca}.\footnote{For that reason, some dilatino helicity amplitudes depend on this reference vector.} The desired Ward identity is obtained by commuting the supercharges $Q(\xi,\bar\xi)$, where $\xi$ and $\bar\xi$ are arbitrary transformation parameters,  with a string of creation operators \cite{Parke:1985pn,Taylor:2017sph}:
\begin{align}0~=~
\big\langle[Q(\xi,\bar\xi)&\, , \,\lambda^- \, g^-\, g^+ \, \phi^*]\big\rangle ~=~[\xi 1]\langle g^- \, g^- \, g^+ \, \phi^* \rangle + \langle \xi 2\rangle \langle \lambda^- \, \lambda^-\, g^+\, \phi^*\rangle \nonumber\\& -[\xi 3] \langle \lambda^- \, g^- \, \lambda^+\, \phi^*\rangle -\langle \xi 4\rangle \langle \lambda^-\, g^- \, g^+ \chi^+\rangle
- m_H \frac{\langle\xi r\rangle}{\langle4r\rangle} \langle \lambda^-\, g^-\,  g^+\, \chi^-\rangle \, , \label{eq:susya3phi*}
\end{align}
After setting $\bar\xi=0,\xi=4$, we obtain
\be
\langle 42 \rangle  \langle \lambda^- \, \lambda^-\, g^+\, \phi^*\rangle = m_H  \langle \lambda^-\, g^-\,  g^+\, \chi^-\rangle \, , \label{eq:susya3phi*step1}
\ee
which relates  $A_3(1^{-1/2},2^{-1/2}, 3^{+1}, \phi^*)$ to an amplitude that can be connected to set 1. To that end, we use the supersymmetric Ward identity
\begin{align}
0=\langle [Q(\xi,\bar\xi)&\, , \, g^-\, g^- \, g^+ \, \chi^-]\rangle=\langle \xi1\rangle \langle \lambda^- \, g^- \, g^+ \, \chi^-\rangle +\langle \xi 2\rangle \langle g^- \, \lambda^- \, g^+ \, \chi^-\rangle\\ & -[\xi3] \langle g^-\, g^- \, \lambda^+ \, \chi^-\rangle + \langle \xi 4\rangle \langle g^-\, g^- \, g^+ \, \phi\rangle + m_H \frac{\langle \xi r\rangle}{\langle 4r\rangle} \langle g^-\, g^-\, g^+ \, \phi^*\rangle \, .
\end{align}
After setting $\bar\xi=0$, $\xi=r=2$, we relate the dilatino amplitude to the MHV amplitude:
\be\langle 21\rangle \langle \lambda^- \, g^- \, g^+ \, \chi^-\rangle = \langle 42\rangle  \langle g^-\, g^- \, g^+ \, \phi\rangle \, .
\ee
By combining this with Eq.(\ref{eq:susya3phi*step1}), we obtain
\be A_3(1^{-1/2},2^{-1/2}, 3^{+1}, \phi^*)=-\frac{m_H}{\langle 12\rangle}A_3(1^{-1},2^{-1}, 3^{+1}\phi)=-\frac{m_H}{\Lambda} \frac{\langle 12 \rangle^2}{\langle 23\rangle \langle 31\rangle},\ee
in agreement with Eq.(\ref{eq:MHV2gim}). Note that only the holomorphic part of supersymmetry transformations (with $\bar\xi=0$) was used in deriving this relation. We conclude that sets 1 and 2 are related by spacetime supersymmetry. Hence celestial supersymmetry appears as a consequence of the holomorphic part of four-dimensional supersymmetry.

\section{Conclusions} In this work,
we discussed supersymmetric Yang-Mills theory coupled to the dilaton field, in the framework of celestial holography.
Previously in Ref.\cite{Stieberger:2022zyk}, we showed that
in the presence of a dilaton background field produced by a pointlike source, celestial gluon amplitudes become well-defined correlators of the products of holomorphic current operators with integer dimensions times the exponential ``light'' operators associated with Liouville theory in the limit of infinite central charge.
In the present work, we discussed the amplitudes involving gauginos and dilatinos, related to MHV amplitudes by supersymmetric Ward identities. We constructed the CCFT operators associated with gauginos
in a similar, factorized form, with the helicity and gauge degrees of freedom contained in the holomorphic supercurrent factors. We showed that in a theory with massive dilaton, the currents and supercurrents form supermultiplets of (1,0) supersymmetry, provided that the R-symmetry breaking amplitudes contribute in the massless limit. This requires a careful choice of
the normalization factors for the operators associated with the gauge supermultiplet. Celestial supersymmetry appears then as a consequence of the holomorphic part of four-dimensional supersymmetry.

The emerging picture is quite similar to heterotic superstring theory. The holomorphic sector of supersymmetric celestial Liouville theory  consists of two-dimensional (1,0) supercurrents carrying all information about the spin and gauge degrees of freedom of the gauge supermultiplet. The role of the Liouville operators is to supplement their integer dimensions to continous values, thus carrying over the information about energies, i.e.\ the scattering data, from spacetime to celestial CFT. There is no supersymmetry in the Liouville sector.

In this work and in the previous Ref.\cite{Stieberger:2022zyk}, we considered a point-like  source creating a dilaton  shockwave propagating on the lightcone.
 The most important question is whether there is some underlying principle that singles out such point-like sources (akin to  ``dilaton background charges'') leading to a holographic description of four-dimensional spacetime in terms of super WZW and Liouville theories. In general, the role of background fields in celestial holography needs further study.

\section*{Acknowledgements}
We would like to thank Tim Adamo, Wei Bu, Davide Gaiotto, Yangrui Hu, and Sabrina Pasterski for useful conversations.
TRT is supported by the National Science Foundation
under Grants Number PHY--1913328 and PHY--2209903, and by the NAWA grant
``Celestial Holography of Fundamental Interactions.''
Any opinions, findings, and conclusions or recommendations
expressed in this material are those of the authors and do not necessarily
reflect the views of the National Science Foundation.
BZ is supported by the Celestial Holography Initiative at the Perimeter Institute for Theoretical Physics. Research at the Perimeter Institute is supported by the Government of Canada through the Department of Innovation, Science and Industry Canada and by the Province of Ontario through the Ministry of Colleges and Universities.

\end{document}